\documentclass[11pt]{elsarticle}
\usepackage{graphics,graphicx}
\journal{Physica A}
\usepackage[margin=2.5cm]{geometry}
\usepackage{xcolor}
\usepackage[linesnumbered,ruled,vlined]{algorithm2e}
\usepackage{algorithmicx}
\usepackage{graphics,graphicx}
 
\begin{document}

\title{Parallel Minority Game and it's application in movement optimization during an epidemic}

\author[label1]{Soumyajyoti Biswas} 
\ead{soumyajyoti.b@srmap.edu.in}
\author[label2]{Amit Kr Mandal}
\ead{amitmandal.nitdgp@gmail.com}
\address[label1]{Department of Physics, SRM University - AP, Andhra Pradesh - 522502, India.}
\address[label2]{Department of Computer Science and Engineering, SRM University - AP, Andhra Pradesh - 522502, India}

\date{\today}

\begin{abstract}
We introduce a version of the Minority Game where the total number of available choices is $D>2$, but the agents only have two available choices 
to switch. For all agents at an instant in any given choice, therefore, the other choice is distributed between the remaining $D-1$ options.
 This brings in the added complexity in reaching a state with the maximum resource utilization, in the sense that the game is essentially
a set of MG that are coupled and played in parallel. We show that a stochastic strategy, used in the MG, works well here too. 
We discuss the limits in which the model reduces to other known models. Finally, we study an application of the model in the context of 
population movement between various states within a country during an ongoing epidemic.    
 we show that   
the total infected population in the country could be as low as that achieved with a complete stoppage of inter-region movements for a prolonged period,
provided that the agents instead follow the above mentioned stochastic strategy for their movement decisions between their two choices. 
The objective for an agent is to stay in the lower infected state between their two choices. We further show that
it is the agents moving once between any two states, following the stochastic strategy, who are less likely to be infected than those not having (or not opting for) such
a movement choice, when the risk of getting infected during the travel is not considered. This shows the incentive for the moving agents to follow
the stochastic strategy. 
\end{abstract}


\maketitle
\section{Introduction}
A large number of agents competing for limited resources is a generic problem appearing in 
a wide range of situations such as in economics, in sharing natural resources to allocating internet bandwidth and so on \cite{phys_rep}. 
There have been many attempts to address these problems from a game theoretic perspective (see e.g., congestion games \cite{ros}).
When the number of choices for an agent is limited to two, the resource allocation problem generally 
falls within the category of the Minority Game (MG) \cite{Challet:1997,Challet:1999,MGBook,Moro:2004}, where an odd number ($N=2M+1$) of agents try to 
be, through repeated and parallely decided autonomous choices, in the minority group, thereby winning a higher share of the
(conserved) resources, or in the language of game theory, a positive pay-off. Clearly, the Nash
equilibrium state for the problem is when the populations $P_1$ and $P_2$ in the two choices are $M$ and $M+1$,
such that one agent cannot benefit by their own action alone provided that the others stick to their respective choices. 
The objective is to reach this state
of minimal fluctuation $\Delta=|P_1-P_2|$  between the two choices in the least possible amount of time, through autonomous and parallel 
decisions by the agents. 

While a complete random decision would keep the population difference between the two choices very high ($\Delta \sim \sqrt{N}$),
 a deterministic cooperative learning mechanism with the past memory of winning choices can reduce it by a constant factor \cite{MGBook}. 
A stochastic strategy \cite{Dhar:2011}, on the other hand, can reduce the fluctuation to the minimum possible value ($\Delta \sim 1$) in a very short time
($t_{sat} \sim \log \log N$). But this strategy requires the additional input of the fluctuation in the previous step to be supplied to the
agents. However, the stochastic
strategy still performs quite well ($t_{sat} \sim \log N$) when $\Delta(t)$ is not exactly know to the agents but a guess value is 
supplied through an appropriate annealing schedule \cite{biswas}.   

However, the total number of choices in a problem is not always limited to two.  Indeed, a general resource allocation problem with multi-choice and multi-agent,
not necessarily of equal number, can be posed \cite{kpr}. The utility factor i.e. the efficiency with which the agents get distributed among the available choices
with least overcrowding, then depends upon the crowd-avoiding switching strategies of the agents \cite{kpr2}. It was shown that if the number of choices and the number of agents
competing for those choices are equal, and an agent is `satisfied' when a choice is occupied by only that agent, the maximum utility fraction for a repeated, 
autonomous choice game is about 0.8 \cite{kpr2}. 
This type of multi-choice, multi-agent game falls under the purview of the Kolkata-Paise-Restaurant (KPR) problem, and have been widely studied (see Ref. \cite{phys_rep} for a review). 
Interestingly, the same stochastic, crowd avoiding strategy works well in both KPR and MG for reaching the maximum possible utility factor. 

\begin{figure}[tbh]
\centering
\includegraphics[width=8cm, keepaspectratio]{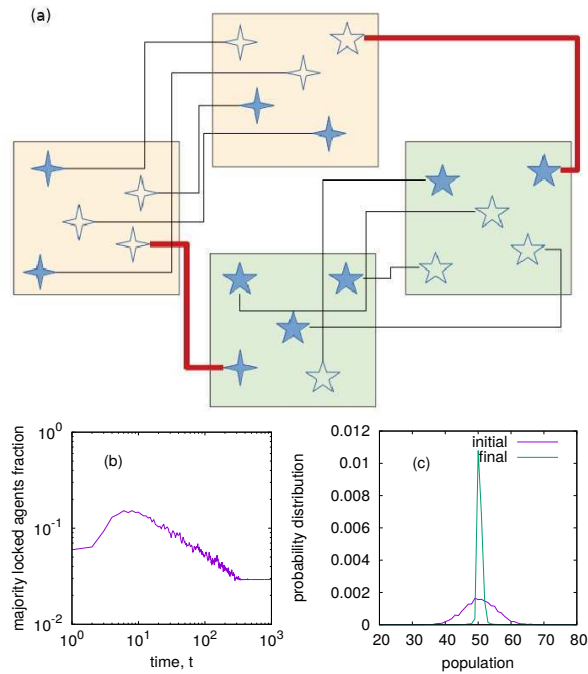} 
\caption{Representation of agents' movement as parallel Minority Games and their effect on epidemic spreading. 
(a) The schematic diagram for the agents with choices in four regions/states. A filled shape represents
a site occupied by an agent and a similar empty shape, connected by a black line, represents their corresponding alternative choice.
Here all agents having an alternative choice are shown. If only similar agents and their alternative choices exist between 
a pair of regions, say the yellow or green regions, then the problem reduces to a pair of decoupled Minority Games.
(b) The time evolution of the fraction of majority locked agents, (c) The initial and final distributions of population. The final distribution shows sharp peaks around $M$ and $M+1$.
}
\label{schematic}
\end{figure} 

In this work, we consider a situation where the total number of choices is $D>2$, but each agent has the option of switching only between two of 
those choices (see Fig. \ref{schematic}).  
The objective for the agents, like in the MG, is to be in the minority of their two available choices. However, the agents
in any given choice, have their alternative choices distributed uniformly between the rest $D-1$ choices, giving an added complexity to the model in the sense that it acts as a set of parallel and coupled MG. 
Such situations are possible in the cases where
the agents have a rank-dependent choice \cite{kpr2} in the limit such that only two choices are viable, e.g., two particular stocks of interests, two 
preferred restaurants among many, or two possible places of residence etc., for a given agent. 

\begin{figure}[tbh]
\centering
\includegraphics[width=8cm, keepaspectratio]{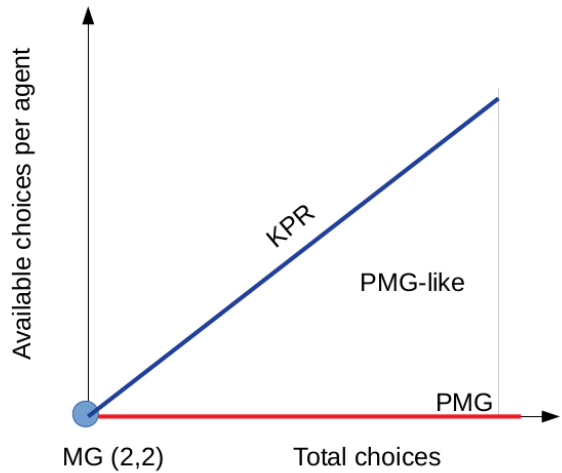} 
\caption{The connection of the parallel MG is shown with other known limits (MG \cite{Challet:1997} and KPR \cite{kpr}).}
\label{games}
\end{figure} 
In particular, we consider two situations, one in which the efficiency of a stochastic switching strategy is studied for the parallel Minority Game
(PMG). Subsequently, we consider an application of the model in terms of the movement optimizations of the agents between their (limited) choice of
residence in a country during an ongoing epidemic. In the second case, the crowd avoidance is to be interpreted in terms of avoiding higher numbers of infected people
in a choice (say, a state within a country). In this case, given that the number of infected people is a function of time, an equilibrium is not defined. However, we study the 
optimized movement strategies in which the risk of infection in the total population and also that of the group opting for switching of residence, are minimized.    

\section{Parallel Minority Game}
Consider a population of $N$ agents and $D$ choices. Assume that a fraction $g$ of the agents have the option of moving between two of those choices (see Fig. \ref{schematic}). 
The other agents do not have any option to switch.
The agents must decide their repeated choices simultaneously at each instant of time. 
The objective for each agent is to be in the minority group
between those two assigned choices. In this section, we look at the case of $g=1$ i.e., all agents have a choice to switch. The case of $g<1$ i.e., the situation where a fraction of the population is immobile, will be considered in the later sections. 

As mentioned before, in the limit $D=2$, the problem is identical to the MG. Also, in a special case, the problem
reduces to $D/2$ decoupled MGs, if the agents in a pair of choices are completely non-overlapping with the agents of any other pair of
choices. However, we look at the case where $D>2$ and the choices can be overlapping. 

Before moving on to the results, it is useful to note the connection of this game with other known limits. In Fig. \ref{games}, we show a schematic
diagram where the total number of choices and the number of available choices to each agent are drawn. When both of these numbers are 2, we get the usual Minority Game. When both these numbers are equal but are greater than 2, then it is the 
multi-choice, multi-agent game, known as the KPR problem. We are interested in the limit where the number of available choices for each agent remain 2, 
but the total number of choices is greater than 2 (parallel Minority Game). In all other cases where the number of available choices for an agent is less than the total number of
available choices, the added complexity due to the overlapping choices of the agents exist. However, in this work we restrict ourselves to two available choices for
each agent.

\subsection{Resource utilization in the model with a stochastic strategy}
The most efficient resource utilization (most number of agents winning a positive pay-off) in the case of the Minority Game is achieved when the 
difference of population between the two
choices is $|P_1-P_2|=\Delta=1$, given that the total number of agents is odd ($N=2M+1$). Similarly, in our case, we can consider a total
population $N=(2M+1)\frac{D}{2}$, with $D$ even, such that a globally efficient resource utilization is the one where $D/2$ choices
have population $M$ and the other $D/2$ choices have population $M+1$. However, due to the distributed choices of the agents, it might happen that 
even for the globally optimized configuration, both options for a given agent are global majority i.e. the populations in both the options for that agent are 
greater than $M$. We call such agents as the `majority locked' agents. Clearly, even with the globally optimized state, those agents never have a chance to 
win a positive pay-off. A proper measure for resource utilization in this case, therefore, is the minimization of the fraction of majority-locked agents after the global
utilization condition is fulfilled. 

\subsection{Stochastic strategy for switching}
As mentioned before, a crowd avoiding stochastic strategy performs best for the MG problem \cite{Dhar:2011}. Specifically, if $P_1$ and $P_2$ are the populations in the 
two choices of the MG, then an optimal distribution of population would be $M$ and $M+1$, where the total population is $2M+1=P_1+P_2$. Clearly, a shift of
$(\Delta-1)/2$ agents would achieve such a distribution, where $\Delta=|P_1-P_2|$. Under a stochastic dynamics, such a shift in the population on average
is obtained when the agents in the majority switches with a probability $p_+=\frac{(\Delta-1)/2}{P_M}$, where $P_M$ is the population in the majority and the agents in the minority do not switch.
The stochasticity would induce a fluctuation of the order of $\sqrt{\Delta(t)}$, which by definition is $\Delta(t+1)$. This gives the recurrence relation
$\Delta(t+1)\approx\sqrt{\Delta(t)}$, giving a very fast convergence (scaling as $\log \log N$, where $N=2M+1$) of $\Delta\approx 1$. The only additional information
supplied to the agents is value of $\Delta(t)$, while traditionally only the sign of it is supplied. 

We proceed with a similar approach for the parallel MG as well. Specifically, we keep the switching probability for a given ($\alpha$-th) agent
 in the majority at a given instant to be 
\begin{equation}
p_+(t)=\frac{(P_i^{\alpha}-P_j^{\alpha}-1)/2}{gP_i^{\alpha}/(D-1)},
\label{mm_str} 
\end{equation}
where $P_i^{\alpha}$ is the population at the 
$i$-th choice of the $\alpha$-th agent, $P_j^{\alpha}$ is the population at 
the location of the other choice assigned to the agent $\alpha$, $g$ is the fraction of agents having the choice of movement 
and $P_i^{\alpha}(t)>P_j^{\alpha}(t)$. The switching probability is zero otherwise. The logic of choosing the particular form of switching probability is the same as before i.e. 
the movement on average will balance the populations between the two states considered at each movement possibility. 

Other than the stochastic strategy mentioned above, we put another restriction in the dynamics that the agents can switch only once during the entire dynamics. This restriction allows for a wider  sample of the agents to move and is particularly useful for the application considered later. 
We ensure, while assigning the choices to the agents, that an equilibrium state exists where both choices of
every agent can have populations $M$ and $M+1$.
  
In Fig. \ref{schematic}(b) we plot the time variation of the fraction of majority locked agents. It saturates to a finite value. For computational simplicity we keep $D=4$. For higher values of $D$, the saturation value is higher. In Fig. \ref{schematic}(c), the initial and final distributions of the populations at various choices are shown. The initial distribution, due to random assignment of the agents in their two choices, is Gaussian. The final distribution is sharply peaked at $M=50$ and $M+1=51$, implying that although the fraction of majority locked agents remain finite, the globally optimized state is nevertheless is nearly reached.

\begin{figure*}[tbh]
\centering
\includegraphics[width=8cm, keepaspectratio]{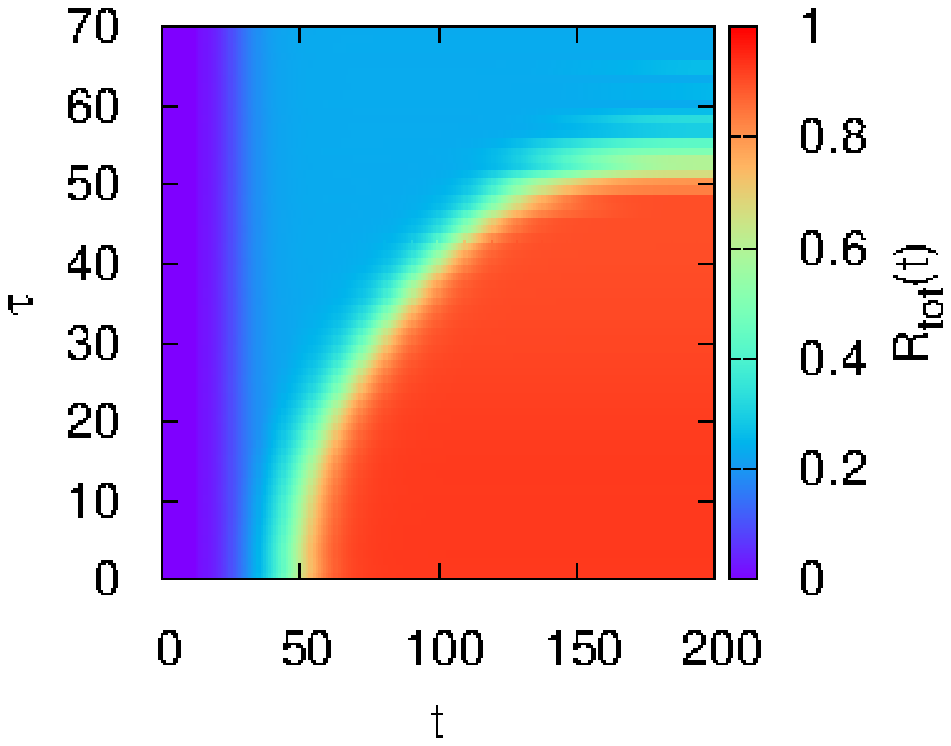} 
\includegraphics[width=8cm, keepaspectratio]{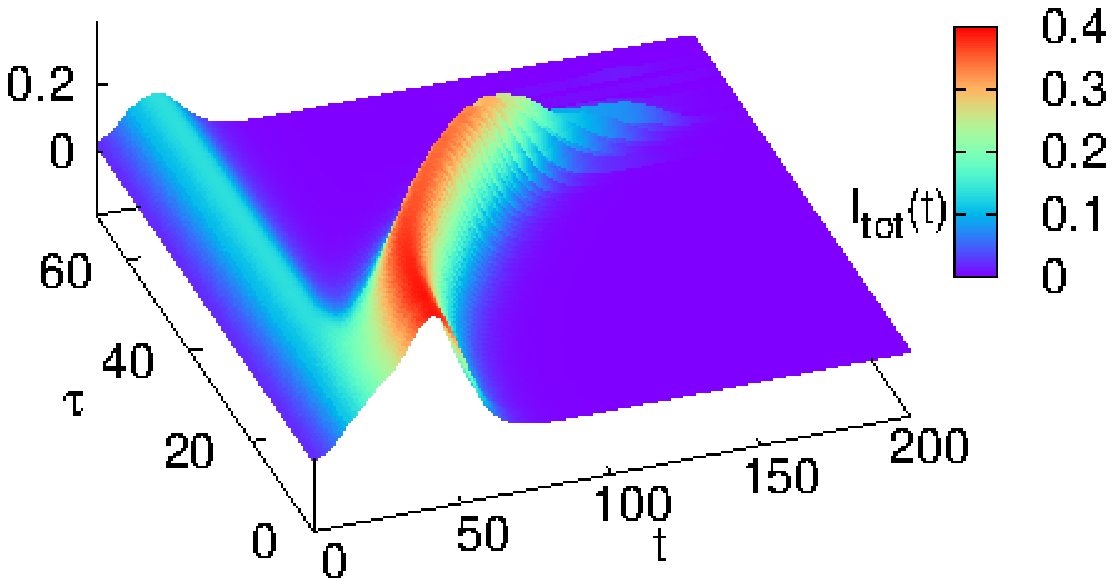}
\caption{The infection spreading in the mean field limit. (a) The total recovered fractions are shown with time for different duration ($\tau$) of lock downs. The 
saturation value of the recovered fraction, representing the agents infected at some point of time, decreases with
prolonged lock down. When allowed, movements always follow Eq. (\ref{move_prob}). (b) The time variation of the infected 
fraction is shown for various duration of lock downs, imposed at $t=0$. A second, more prominent peak is observed,
unless the lock down period is long ($\sim 60$ steps).}
\label{kpr-pmg}
\end{figure*} 

\begin{figure}[tbh]
\centering
\includegraphics[width=13cm, keepaspectratio]{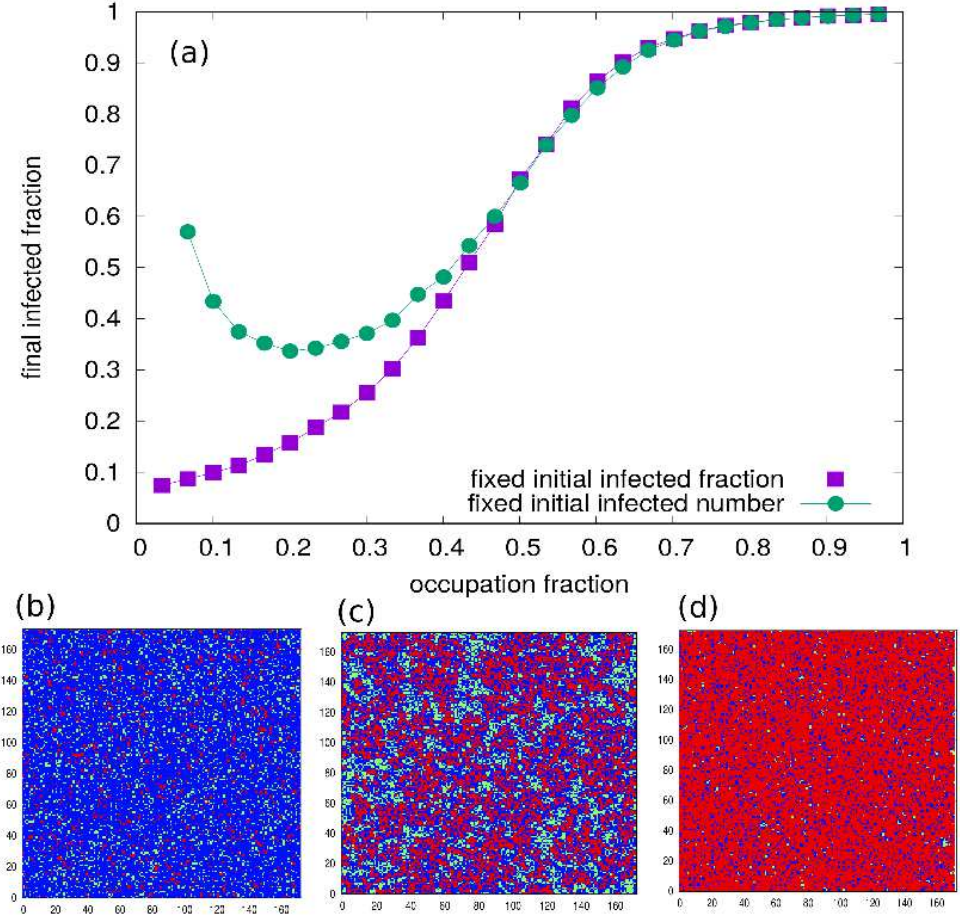} 
\caption{SIR model on diluted square lattice. (a) The total infected fraction is plotted with the variation of occupation fraction. The simulation is done only for one region with size $173 \time 173$,
hence no movement strategy is present. The two curves represent the situations where the initial infection comes from a fixed number of infected agents ($1000$),
 or a fixed fraction of infected agents ($1/15$). The three configurations show the susceptible (green), recovered (red) and vacant sites (blue) at the end of the simulation,
for occupation fractions $0.167$ (b), $0.501$ (c) and  $0.835$ (d), for a fixed number (1000) of infected agents at $t=0$. For the simulations in the text, the half occupancy configuration is used.}
\label{perco}
\end{figure} 

\begin{figure}[tbh]
\centering
\includegraphics[width=15cm, keepaspectratio]{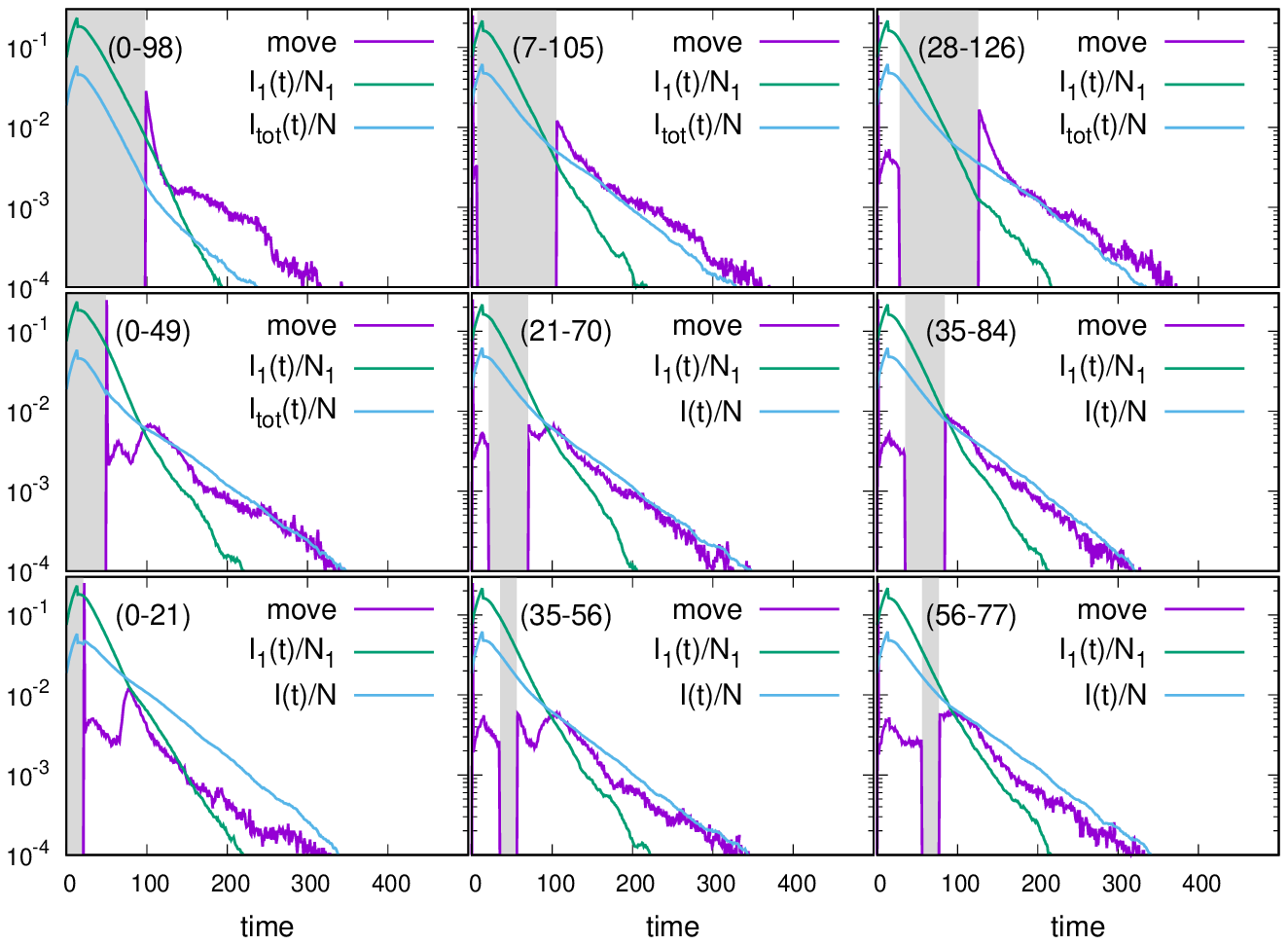} 
\caption{The time evolution of the total fraction of agents moved, the infected fraction in $D1$ and the total infected fraction, each
independently normalized (see text) and plotted for various duration and start times of lock down, for the model implemented in a square 
lattice. The lock down start and end times are indicated
in each figure. Inevitably, a spike in the movement is noted immediately following the lifting of lock down. This almost in all cases results
in a slowing down of the total recovery rate.}
\label{inf_and_move}
\end{figure} 

Therefore, we see that in PMG the stochastic strategy in Eq. (\ref{mm_str}) along with the restriction of one movement per agent
gives a distribution of population such that the globally optimal solution is nearly reached ($D/2$ choices have population $M$ and $D/2$ choices have population $M+1$),
along with a low value for the number of majority locked agents.
We now proceed with the particular application of the game with this strategy.

\section{Movement optimization during an epidemic using the parallel Minority Game}
As mentioned before, there are a number of situations where the parallel Minority Game can be applied. The game is essentially a limit of the multi-choice games
where the choices are rank dependent to the extent that only two choices are viable for an individual agent.
Here we consider a particular application of the parallel Minority Game in terms of optimizing the movement of population during an ongoing epidemic, with the objective of minimizing the total infection. Optimization of agent movement to avoid infection have been studied before. For example, in Ref. \cite{yang18,yang16} the agents migrate out of a location if the infection rate there is above a threshold. But in our case, the choices of movements are not only dependent on the local environment but also on the environment of the location to which the agent can move. 

During the ongoing global crisis since the outbreak of the COVID-19 pandemic \cite{who}, many countries around
the world have imposed varying degrees of restrictions on international as well as their domestic travels \cite{chin}. 
At times when such restrictions are fully or partially lifted,
an inevitable question for a significant fraction of the population would be to change their 
locations in order to escape the regions of higher
infections. But due to difficulties in detecting the disease, an influx of population to relatively
less infected regions can increase the infection rate in those regions, making them less viable options in the subsequent times.

If
a country has $D$ states/regions with different rates of infection, one person may have a rather limited option in terms of 
the number of regions in which they can live for a long time (e.g., the state of their work place and the state of their hometown).
Indeed, a small fraction of the total population would enjoy such a choice \cite{migrants}, meaning the $g<1$ in this case.
 Clearly, in a given region, different persons will have their
choices/options distributed among the rest of the $D-1$ regions i.e., mirroring the situation described in the parallel Minority Game in the previous section. 
The objective of the agents is to be in the minority group in terms of the number of infected people. 
In the usual Minority Games, the variations in the target variable (i.e., populations in the two choices), take place due to
the switching of choices of the agents. In this case, the target variable is the number of infected people. In a given region, it can change due to (a) movements of agents from
the other $D-1$ regions/states and (b) due to the evolution of the epidemic within the region itself.  Given that the target variable for minimizing fluctuation (difference in the infected population) is not a conserved quantity, an equilibrium state is no longer defined. However, a well defined quantity is the total infected population in the country 
after the end of the epidemic. 
\begin{figure}[tbh]
\centering
\includegraphics[width=15cm, keepaspectratio]{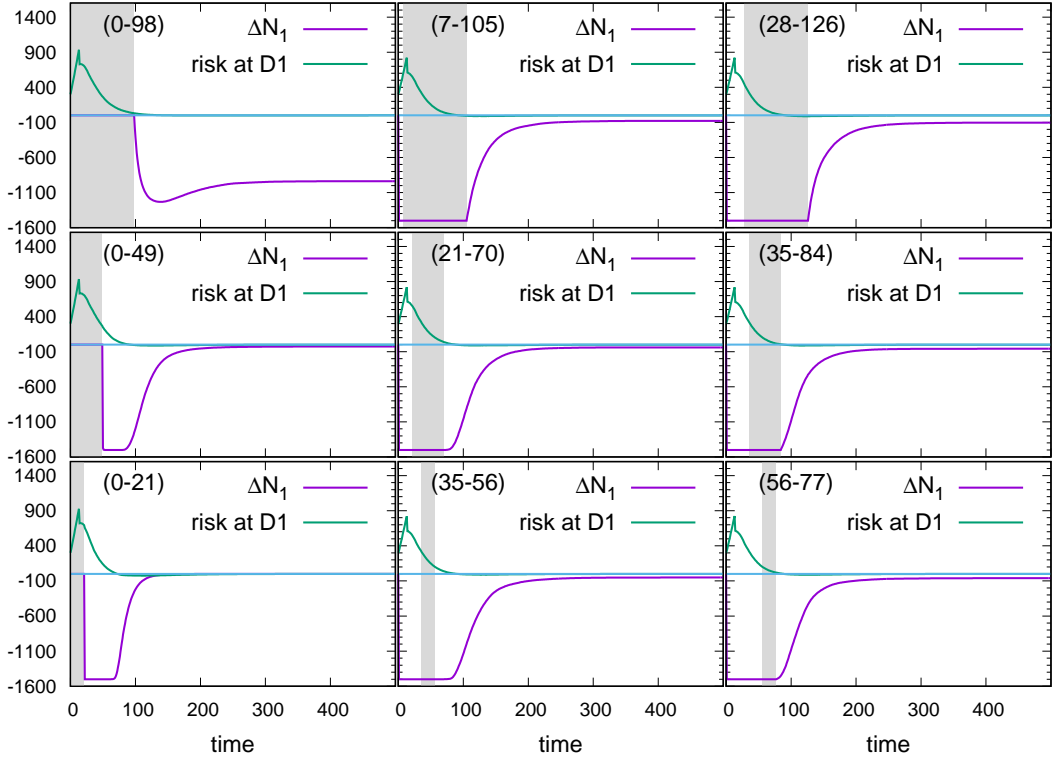} 
\caption{We plot the time evolution of the relative risk at $D1$ with respect to the average risk in the remaining regions i.e., the difference between the infected fraction at $D1$ and the average of the infection fraction in the remaining regions. This quantity is multiplied by an arbitrary factor (400), for easier visualization. The change in the population in $D1$,
is anti-correlated with the risk (see text for details).}
\label{risk}
\end{figure}

\begin{figure}[tbh]
\centering
\includegraphics[width=15cm, keepaspectratio]{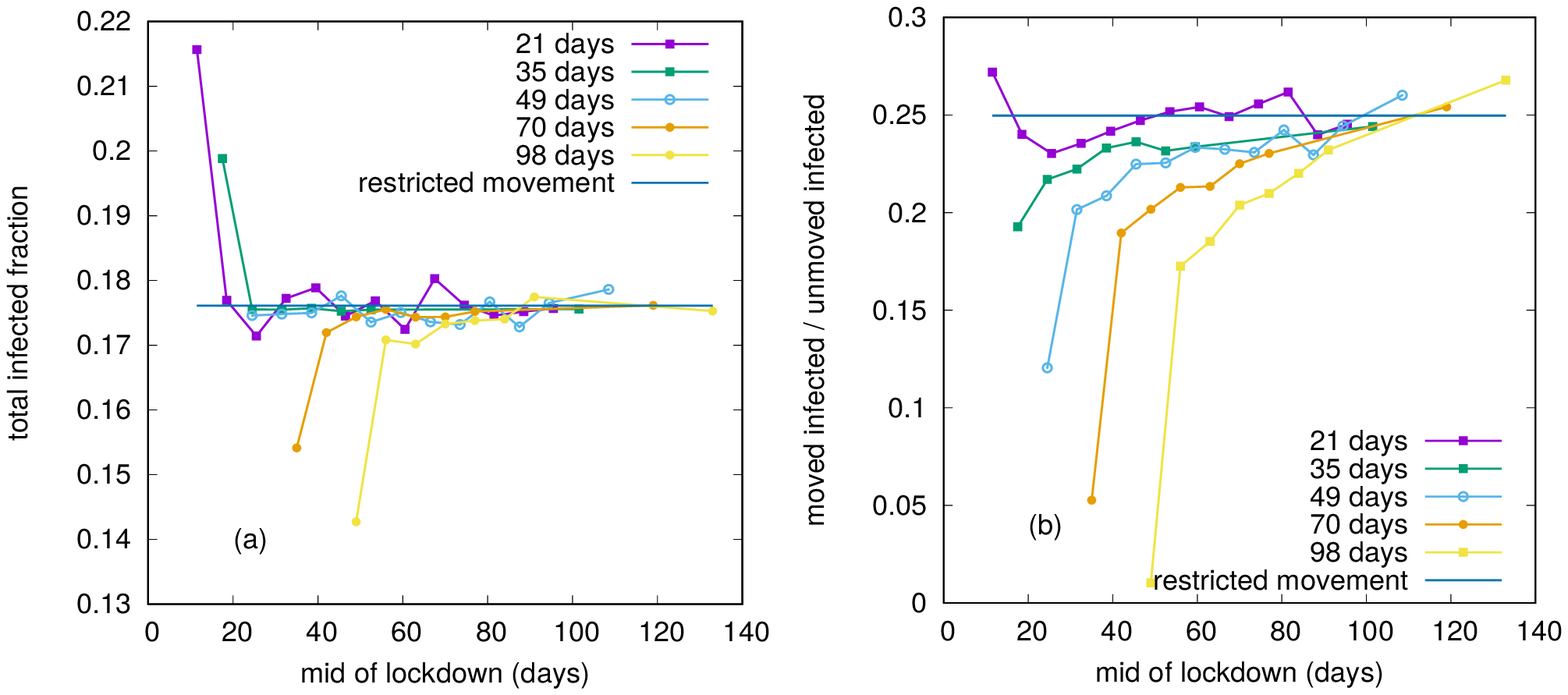}
\caption{Effect of the stochastic strategy driven movements in epidemic spreading. (a) The fraction of the total infected population, at the end of the epidemic dynamics, 
is shown for various duration of lock downs against the mid-point of the lock down period for the model in th square lattice.
 A small initial lock down is least effective.
Comparisons are made with the infected fraction for no imposition of inter-state movement, shown in the horizontal line. This seems 
to work better than a small period lock down, or equally good as the late imposition of long lock down. The movement strategy, in all the cases, 
follow Eq. (\ref{move_prob}). If the movements are random (agents switch with $50\%$ probability), the infected fraction is $0.197$, which is
worse than almost all the cases.  
(b) The relative infection probabilities are shown for the agents opting for switching their
location divided by the corresponding probability for the agents who did not switch. Given the ratio is almost always less than 1, 
the population opting for the switching following Eq. (\ref{move_prob}) are less likely to be infected, where we did not consider the
chance of getting infected during the travel.}
\label{fig3}
\end{figure} 

There is a long history of mathematical models for epidemic spreading \cite{ep1,ep2,ep3}.
There has already  been a multitude of attempts in simulating the COVID-19 pandemic with varying degree of 
realistic features and also analysis of the available data (see e.g., \cite{covid1, covid2, covid3, covid4, covid5, covid6}).

However, our aim here is to look for a generic strategy for optimized movement between various regions for some of the agents to avoid infection. 
Therefore we keep the epidemic spreading
 to be a standard  Susceptible (S)- Infected (I)- Recovered (R) model \cite{sir_o}  type model. At an instant, 
one susceptible agent can get infection, with a probability $\beta$, from any neighboring infected agent.  An infected agent remains infected for a 
period $\tau=14$ days and then recovers. A recovered agent is immune to further infection. 
As an example, given that the reproduction rate $R_0$ for the COVID-19 pandemic is about $2.28$ \cite{pubmed}, we keep the infection probability
$\beta=2.28/(\tau z)$, as $\tau=14$ and $z$ depends on the topology of the underlying lattice i.e., the connectivity among the agents. 
While this gives a correct order for the infection probability, this is by no means a precise estimate of the same, which will 
vary due to the clustering of the locations, immunity variations of the agents and so on.

A reasonable restriction for the movement is that one agent can switch only once during the entire period.
For the agents who can switch, we apply a similar stochastic strategy as in Eq. (\ref{move_prob}) i.e., one agent ($\alpha$) in a
higher infected region (say, $i$-th region) will move with a probability
\begin{equation}
p_+=\frac{(I_i^{\alpha}-I_j^{\alpha})/2}{gP_i^{\alpha}/(D-1)},
\label{move_prob}
\end{equation} 
where $I_i^{\alpha}$ and $I_j^{\alpha}$ are the numbers of infected populations in the two choices ($i$-th and $j$-th) of the $\alpha$-th agent, $P_i^{\alpha}$ and $P_j^{\alpha}$ are the respective total populations 
and $I_i^{\alpha}(t)>I_j^{\alpha}(t)$. The agents already in the lesser infected region compared to their other alternative choice
 at an instant do not switch at that instant ($p_-=0$).
As before, the factor $(D-1)$ is kept as the agents can move to any one of the $D-1$ regions, giving an average shift of $(I_i^{\alpha}-I_j^{\alpha})/2$ to the $j$-th region, which
balances the infections in the two regions. 
Note that as the infected number is not conserved, the total number 
may not be odd, therefore we remove the $-1$ factor in the numerator, as used in Eq. (\ref{move_prob}).

At each time step, first the SIR model is implemented for all agents. Then the movements are made 
following Eq. (\ref{move_prob}). A parallel update rule is followed, such that the changes in both the epidemic part and the movement part, are reflected
in the following time step (see Appendix for the algorithm and Ref. \cite{code} for the code used). Furthermore, we study the effect of a lock down period, during which
all inter-state movements are stopped. During the lock down period, the SIR model runs, but the movements do not happen.

Below we consider two topologies for simulating the model. First we consider the simplest case i.e. the mean field limit. Then we consider the case which is somewhat
more realistic i.e., a sparsely populated square lattice. 

\subsection{Results in the Mean field limit}
In the simplest case where a parallel MG is defined is when $D=4$. We begin with a mean field version with $D=4$, where each agent in any region can 
interact with $z=4$ randomly selected agents in the same region. There is no interaction between agents belonging to two different regions 
at a given time.  As indicated before, not every agent has the choice of relocating to a different state for a long time. We assign  a fraction $g=0.1$ \cite{migrants} 
of the agents with a choice 
to move to one of the remaining 3 locations only once during entire period of the epidemic dynamics. The remaining question is then the decisions of the agents
with movement choices to switch to their alternative locations, which is done following Eq. (\ref{move_prob}) during the times when a lock down (stoppage of inter-state movements)
is not in place.

\begin{figure*}[tbh]
\centering
\includegraphics[width=8cm, keepaspectratio]{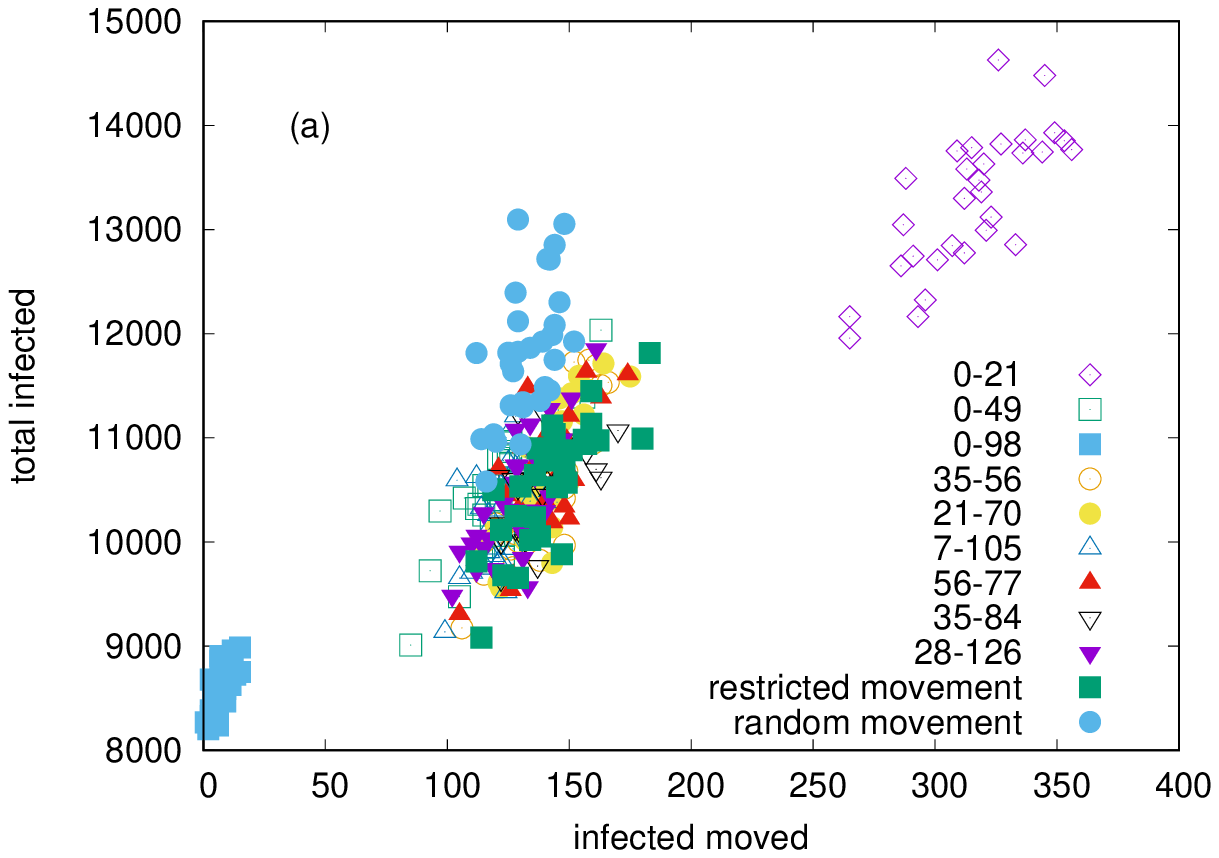}
 \includegraphics[width=8cm, keepaspectratio]{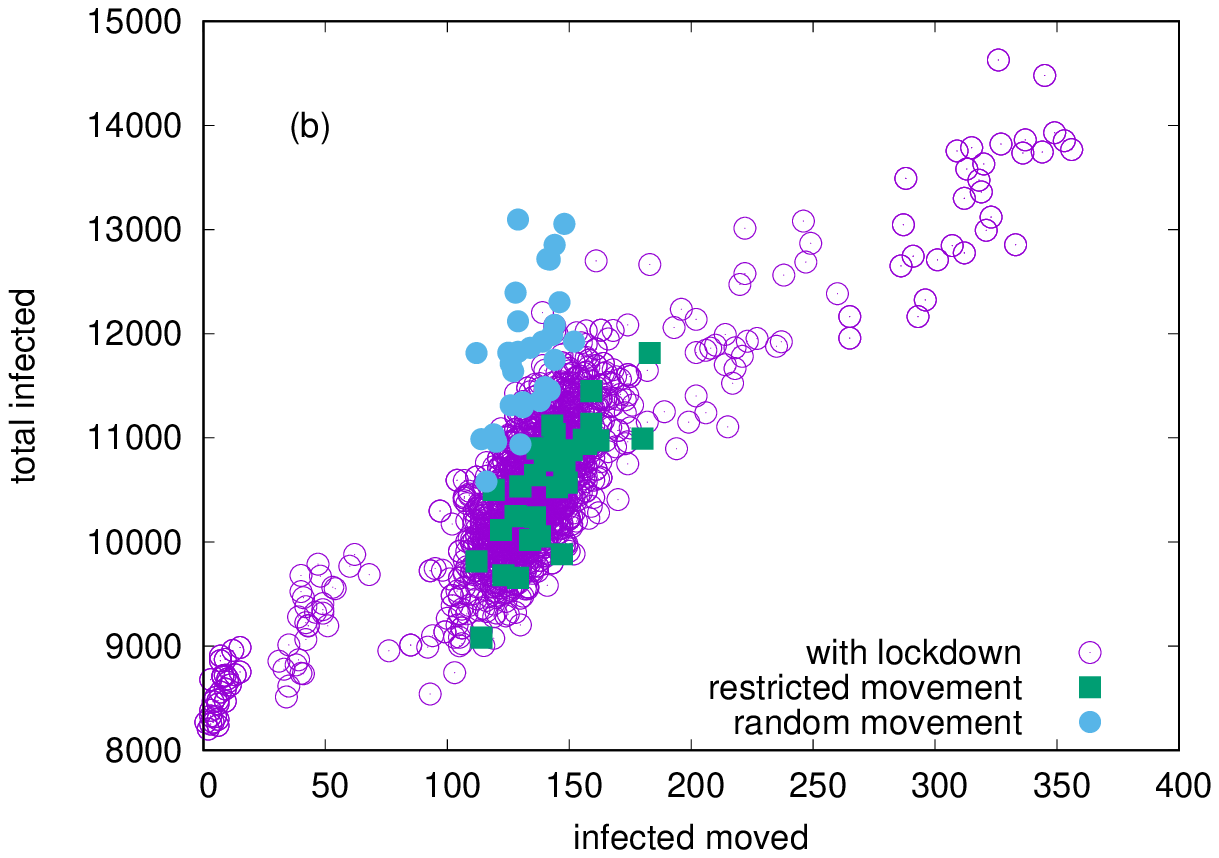}
\caption{(a) The variation in the final infected number is shown with the number of infected agents movement. The total infected number depends upon
the total infected agents moved, irrespective of the lock down periods or even just the movement with stochastic strategy (restricted movement). However, if the
stochastic strategy is not followed, and the movements are random (maximum movement per agent is still 1), then even for the similar values of infected agents moved, the total 
infection is higher. This shows the utility of the stochastic strategy. (b) The same quantities are shown for the three classes of movements, stochastic strategy with lock down, stochastic strategy without lock down and random strategy without lock down.}
\label{m_i}
\end{figure*} 

 Fig. \ref{kpr-pmg}(a) shows the 
variation in the total recovered fraction with time, for various duration of lock down, starting at $t=0$. At $t=0$, all agents are in susceptible state, 
except some of the agents in one region who are infected, which act as the starting seed for the epidemic. We consider a population size of $15000$ for each state 
(total population $60000$) and an initial infection for $1000$ agents in one region (say, $D1$). The results are averaged over 1000 initial conditions. Fig. \ref{kpr-pmg}(b) shows the variation of the number of
infected agents with time for various lock down periods ($\tau$) imposed at $t=0$. It shows that a primary peak is always present, which is due to the spreading 
of infected in $D1$. However, an early lifting of lock down prompts the agents in $D1$, who have the choice, to switch to the other regions and thereby
bringing infection to those regions, resulting in a secondary peak, which is often more prominent. A prolonged lock down ($\sim 60$ time steps), 
however, reduces and eventually eliminates the second peak. For a shorter lock down, the total recovered fraction approaches unity in the long time
$R_{tot}(t\to\infty)\to 1$, i.e. almost everyone is infected at some point of time. 
Although the infection probability for the moving agents are comparatively lower (not shown), the benefit of the stochastic strategy is not clear here due to 
the overall high infection rate. 
This is due to the mean field connectivity among the agents within a region, which is rather unrealistic. Therefore, 
we need to consider a less connected topology.   
 
\subsection{Results in two-dimensions}
 To look for a more realistic connection topology such that the total infected fraction is significantly below 1, the simplest choice is a
square lattice ($L\times L$) to represent possible locations for agents in each region.  A susceptible agent can get infected from any one of its
nearest and next nearest (diagonal) infected neighbors. Open boundary conditions are used here; given the dynamic nature of the neighbors due to
movements, this is not expected to affect the results presented below. The results are averaged over 40 initial realizations i.e., different configurations of occupancy, initial infected agents and agents having choice to move.

\subsubsection{Choice of occupation fraction}
 We take $L=173$ and each state is 
initially occupied with $15000$ agents i.e., an occupation fraction of about $0.5$, giving an average coordination number $z=4$. This is
close to the site percolation threshold \cite{stau}, which represents a reasonable restriction in interactions among the populations in terms of either the imposed
social distancing or due to other factors seen in epidemic spreading \cite{ps1,ps2}.
 An occupancy below this would make the population clusters
fragmented, hence having very less chance of infection spreading. A higher occupancy will form a compact structure, with a very high 
infection spreading chance. Indeed, population densities are often empirically found
to form a fractal structure (see e.g., \cite{city}), which is ensured at the percolation transition point. 

To underline the validity of the choice, we simulate the SIR epidemic in a square lattice for various levels of occupancy. This is for one region only, therefore no movement is allowed.  We show in Fig. \ref{perco}(a) the variation of the total infected fraction with various levels of occupancy 
of the lattice, when the initial infection either comes from a fixed number of infected agents ($1000$), or a fixed fraction of infected agents
($1/15$). Figs- \ref{perco}(b)-(d) show snapshots of the lattice for three occupation fractions (0.167, 0.501, 0.835) at the saturation point. It shows that the half-occupancy keeps the total infected fraction well below 1. Therefore, for the subsequent simulations, we keep the model parameters 
as mentioned above. This means that the average coordination number here is also $z=4$, as was done earlier for the mean-field version of the model.

\subsubsection{Effect of infected fraction on agents' movement}
In Fig. \ref{inf_and_move}, we show the time evolution of the number of agents moving (normalized by the total number of agents having such a choice of movement), the fraction of the infected population in $D1$ (normalized by the instantaneous population at that
region $N_1(t)$) where the initial infection started and the total infected fraction of all four regions, for various duration
of lock downs i.e., complete stoppage of inter-state movements and for various start times of such
a lock down. Invariably, a spike in the movement is noticed immediately following the lifting of a lock down, which then subsequently 
decreases. This is intuitively clear, as a lock down will keep the initially infected region highly infectious and immediately
after the lifting of the lock down, agents will move out of it. Almost in all cases, the lifting of the lock down 
slows down the recovery rate in the total infected fraction. This is due to deconfinement and spreading of infections among the
different regions. There is also a secondary peak in the moving fraction when the infection fraction in $D1$ and the total infection fraction becomes equal. This is due to the movements to $D1$ from other regions, as it becomes less infected compared to the other regions.
This point is further elaborated in Fig. \ref{risk}. Here we show the time evolution of `risk' at $D1$, which is the 
difference between the infected fraction at $D1$ and the average infected fraction in the remaining three regions i.e., $\frac{I_1(t)}{N_1(t)}-\frac{I_2(t)+I_3(t)+I_4(t)}{N_2(t)+N_3(t)+N_4(t)}$. We also show
the change in the total population at $D1$ i.e., $\Delta N_1(t)=N_1(t=0)-N_1(t)$. In all cases, the initial risk at $D1$ is high. If there is no lock down, the 
population at $D1$ sharply decreases. In case of initial lock down, the decrease is seen as soon as the lock down is lifted. 
In cases of intermediate lock down periods, the population at $D1$ changes back to the original value following the lowering 
of risk at $D1$, which is what we mentioned above.  

\begin{figure*}[tbh]
\centering
\includegraphics[width=8cm, keepaspectratio]{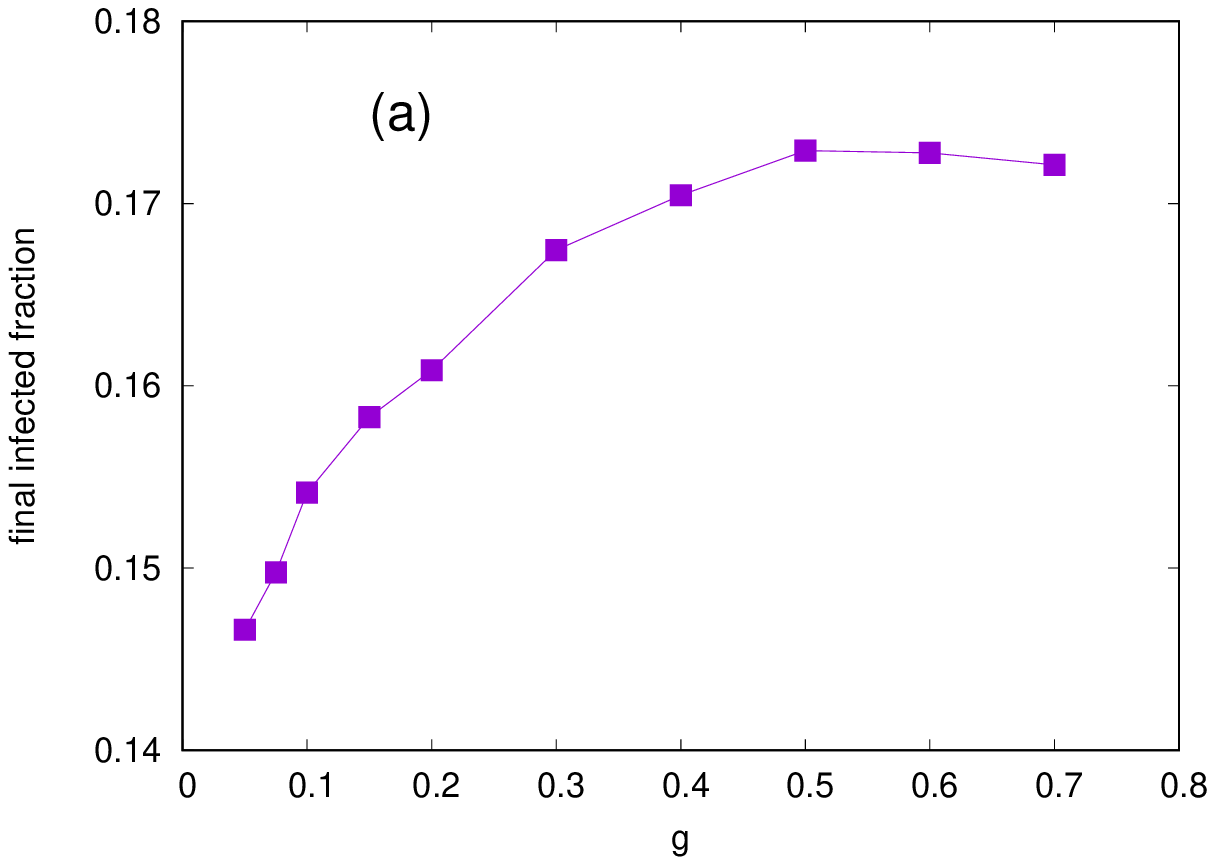}
 \includegraphics[width=8cm, keepaspectratio]{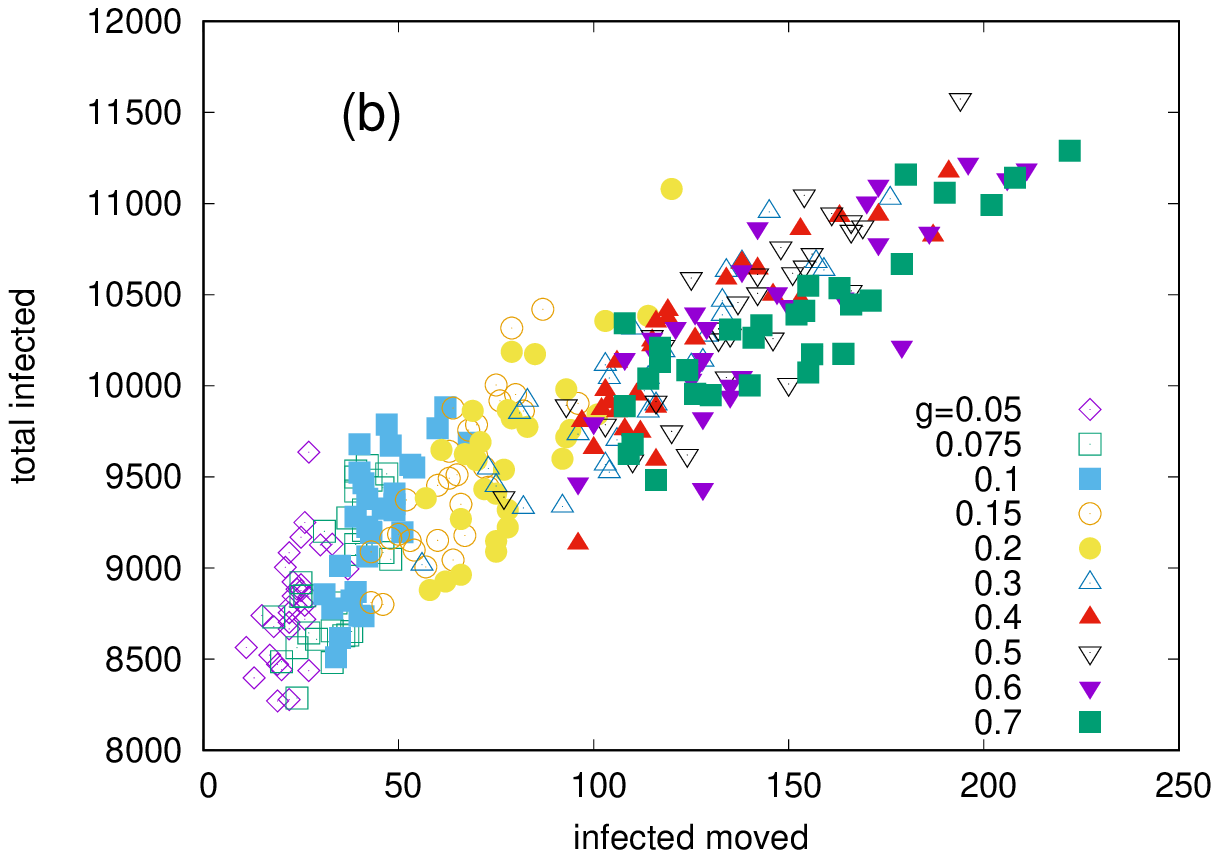}
\caption{The dependence of final infected population on the fraction of agents having a choice to switch. (a) The fraction of agents ($g$) having a choice to switch is plotted against the final total infected population. Initially the infected fraction increases with $g$, but beyond $g\approx 0.5$, the dominant factor in movement restriction is the relative risks among the regions and not the lack of choices for the agents, hence the infected fraction saturates. (b) However, if only the number of infected agents' movement is considered, then an approximate linear relation is seen, similar to what is observed in Fig. \ref{m_i}. These simulations are for lockdown period 0-70 days.}
\label{g_variation}
\end{figure*} 
To quantify the effects of movements on the total infection, in Fig. \ref{fig3}(a) we show (horizontal line) the
 total infected fractions when no stoppage of inter-state travels is imposed. But the restriction that
the agents can switch only once, following 
 Eq. (\ref{move_prob}), remains. We then compare the total infected fraction, at the end of the epidemic dynamics, 
for various duration of lock downs. We see that the strategy of only the restricted movements performs better 
almost in all cases (see Fig. \ref{fig3}(a)) than the restricted movement with a lock down period, unless a 
prolonged lock down ($\sim 60$ time steps) is imposed at $t=0$. 
If the ratio of the infection probability is calculated for the agents 
opting for a switch in their choices following Eq. (\ref{move_prob}) and the same for the agents who did not switch, we see that
the ratio is almost always less than 1, indicating on average a lower risk of infection for the 
agents making the switch (Fig. \ref{fig3}(b)).
The total infected fraction for a random movement (agents having choice shift with 
probability $0.5$ once), is $\approx 0.197$ (not shown), which is higher
than only the restricted movement with stochastic strategy and the most cases of lock downs.
This indicates a stability of the strategy itself, as the agents following it gets benefited, while also benefiting 
(although by a lesser amount) all the remaining agents.

 In Fig. \ref{m_i}(a),
we show the total infected number with the total number of infected agent movement. For different lock down start time and duration, the infected number seem to
depend only upon the infected movement, and not when such movements occurred (or the lock down for that matter). This is true, as long as the stochastic strategy
in Eq. (\ref{move_prob}) is followed. If the movements are random, then the total infection is higher (see Fig. \ref{m_i}(b)) than that following the stochastic strategy, even for a similar
number of infected agent movement. 

On a similar note, if more agents have the choice to move i.e. the value of $g$ is increased, intuitively it is expected that the final infected fraction will also increase with $g$. This is true for small values of $g$, but beyond a threshold value ($\approx 0.5$), the final infected fraction saturates (see Fig. \ref{g_variation}(a)). At that stage, the switching or movement is no longer predominantly affected by lack of choice, but the major factor is the relative risks of infections among the regions. On the other hand, the final infected fraction is still an approximately linear function of the number of infected agents moved (see Fig. \ref{g_variation}(b)), similar to what is seen in Fig. \ref{m_i}. For these results, the lockdown duration is 0-70 days.

\section{Discussions and conclusions}
In this work, we introduce a parallel Minority Game, where $N=\frac{D}{2}(2M+1)$ agents try to be in the minority group through repeated attempts between their two randomly assigned choices but the total number of available choices is $D>2$. Therefore, for the population in any given choice, the second choice is uniformly distributed between the remaining $D-1$ options. In the limits $D=2$, the game reduces to the usual Minority Game, but for $D>2$ it has the added complexity of simultaneously reducing the global fluctuation ($D/2$ choices have population $M$ and $D/2$ choices have population $M+1$) and the fraction of agents for whom both options have population greater then $M$.
We use a stochastic strategy (Eq. (\ref{mm_str})) for the switching of the agents and an added restriction that each agent can switch only once during the dynamics. We find that the both of the above mentioned objectives are nearly satisfied for this strategy. 

We then apply this movement strategy for the population in a country where the agents move among the different states/regions during an ongoing epidemic. However, only a fraction ($g=0.1$) of the population have the choice of movement. The remaining population participate in the epidemic dynamics but do not change their locations. The objective for each mobile agent in this case is to be in the region with lower number of infected people. The epidemic spreading follows a standard SIR dynamics. We simulate the model for both a mean field and a two dimensional square lattice. The qualitative features of the results are same in both cases. 

 It is seen that the total number of infected fraction is lower if the agent follow a similar stochastic strategy (Eq. (\ref{move_prob})) than the case where the agents can switch randomly. This movement strategy also works better than  short and intermediate duration lock downs. This is because, during a lock down (stoppage of all switching activities) the epidemic grows in the region where it started. If the lock down duration is not long enough, the infected population in that region immediately spreads out to other regions, making those regions infected. However, almost in all cases, it is seen that the agents who follow the movement strategy mentioned above are relatively less infected than the agents who do not have such a choice of movement. This is significant in the sense that the mobile agents are not benefited from abandoning the movement strategy (as they are relatively less infected if they follow the strategy). The immobile agents are also benefited if the mobile agents follow this strategy, since a random movement increases overall infected number.

In conclusion, we have introduced a new version of a parallel and coupled Minority Game, with $N$ agents switching between two of the $D$ available choices. We find a stochastic strategy that performs well in terms of resource utilization in this model. We apply the model for the movement strategies for a population during an ongoing epidemic. A similar stochastic strategy is found to keep the total infection lower than that with random movements.

\section*{Appendix}
The algorithm used for the simulations is given below:

	\begin{algorithm}[H]
		\SetKwInput{KwInput}{Input}                
		\SetKwInput{KwOutput}{Output}              
		\DontPrintSemicolon

		\SetKwFunction{FPerson}{Person}
		\SetKwFunction{FGrid}{Grid}
		\SetKwFunction{FRegion}{Region}
		\SetKwFunction{FgetMovementProbability}{getMovementProbability}
		\SetKwFunction{FInitialize}{Initialize}
		\SetKwFunction{FSimulate}{Simulate}
		\SetKwFunction{FAllowMovement}{AllowMovement}
		
			\SetKwProg{Fn}{Def}{:}{}
		\Fn{\FPerson}{
 			state \tcp*{1 = Susceptible, 2 = Infected, 3 = Recovered}
			choice \tcp*{region id where it can move}
			initialRegion \tcp*{starting region of the person}
			stateWhenMoved \tcp*{state of the person during movement} 
			currStateTimeStep\tcp*{tracks infection days}\;
		}

			\SetKwProg{Fn}{Def}{:}{}
		\Fn{\FRegion}{
			regID  \tcp*{ID of the region}
			population \tcp*{total population in the region}
			endTime \tcp*{no of days for the simulation}
			initInfectedPopulation \tcp*{initial Infected Population}
			infectionProbability \tcp*{probability of getting infected}
			infectionDuration \tcp*{duration of infection}
			choiceFraction \tcp*{\% of people can move}
			Grid prevGrid \tcp*{holds the previous state of the region}
			Grid currGrid \tcp*{holds the current state of the region}
			susceptibleCounts \tcp*{Susceptible population in the region}
			infectedCounts \tcp*{Infected population in the region}
			recoverCounts \tcp*{Recovered population in the region} 
		}

			\SetKwProg{Fn}{Def}{:}{}
		\Fn{\FGrid}{
			Person[ ][ ] grid \tcp*{2D array of persons}
			maxRows \tcp*{calculated from population size}
			maxColumns \tcp*{calculated from population size}
		}

		\SetKwProg{Fn}{Function}{:}{}
		\Fn{\FInitialize{ }}{
			determine size of $grids$ from the total $population$ \& percolation point;\; 
			set initial $status$ to $Susceptible: 1$;\;
			randomly populate $prevGrid$ and $CurrGrid$; \;
			randomly place the $initInfectedPopulation$ on the $grids$ by changing its state to infected: 2;\;
		}

	\end{algorithm}
		\pagebreak
		
	\begin{algorithm}[H]
		\SetKwProg{Fn}{Function}{:}{}
		\Fn{\FSimulate{$first$, $second$}}{
			ccurrentTime = 0\;
			\While{currTime != endTime}
			{
				\For { k = 0; k $<$ regionCount; k++} 
				{
					\For {int i = 0; i $<$ $region_K$.MaxRows; i++}
					{
						\For {int j = 0; j $<$ $region_K$.MaxColumns; j++}
						{
							Person $person$ = $prevGrid_k$.getPerson(i,j)\;
							\If{person is Susceptible}
							{
								\For {all the $neighbour$ of a person in prevGrid}
								{
									\If { neighbour is Infected in prevGrid}
									{
										calculate a randomProbability for the person \;
										\If {randomProbability $<$  $region_K$.infectionProbability}
										{
											Marked Person as \textit{Infected} in $currGrid$ \;
										}
									}
								}
							}
							\If {person is Infected}
							{
								\If{currStateTimeStep == infectionDuration}
								{
									Marked Person as \textit{Recovered} in $currGrid$ \;
								}
							}
						}
					}
					Store susceptibleCounts, infectedCounts and recoverCounts \;
					
				}
			   currTime++\;
			   \If{currTime not within lockdown period}
			   {
			   		\FAllowMovement(region, currTime);
				}
			}
		}
		
		\SetKwProg{Fn}{Function}{:}{}
		\Fn{\FAllowMovement{region, currTime}}{
			\For{all possible inter-region (regionX, RegionY) movement}
			{
				moveProb = \FgetMovementProbability(regionX, RegionY)\;
				\For{each person in rgionX}
				{
					\If{have choice to move to regionY}
					{
						calculate a randomProbability for movemet \; 
						\If {randomProbability $<$  $region_{X,Y}$.movementProbability}
						{
							Move that person: regionX $\rightarrow$ RegionY \;
						}
					}
				}
			}
		}
		
		\SetKwProg{Fn}{Function}{:}{}
		\Fn{\FgetMovementProbability{regionX, regionY }}{
			$prob = ((regX.infectedCounts-regY.infectedCounts)/2)/(regx.totalPopulation*(regx.choiceFraction/(TotalNoRegions - 1)))$\;
		}
	\end{algorithm}


\section*{Acknowledgements}
The authors acknowledge Bikas K. Chakrabarti, Parongama Sen and Arnab Chatterjee for their comments and suggestions on the manuscript.


\end{document}